\documentclass[journal=nalefd,manuscript=letter]{achemso}
\usepackage{chemformula} 
\usepackage[T1]{fontenc} 
\mciteErrorOnUnknownfalse
\usepackage{graphicx}
\graphicspath{ {./images/} }
\usepackage{siunitx}

\title{Scanning NV magnetometry of focused-electron-beam-deposited cobalt nanomagnets}

\author{Liza \v{Z}aper} 
\affiliation{Department of Physics, University of Basel, 4056 Basel, Switzerland} 
\alsoaffiliation{Qnami AG, 4132 Muttenz, Switzerland}

\author{Peter Rickhaus} 
\affiliation{Qnami AG, 4132 Muttenz, Switzerland} 

\author{Marcus Wyss} 
\affiliation{Swiss Nanoscience Institute, University of Basel, 4056 Basel, Switzerland} 

\author{Boris Gross} 
\affiliation{Department of Physics, University of Basel, 4056 Basel, Switzerland} 

\author{Martino Poggio} 
\affiliation{Department of Physics, University of Basel, 4056 Basel, Switzerland} 
\alsoaffiliation{Swiss Nanoscience Institute, University of Basel, 4056 Basel, Switzerland}

\author{Floris Braakman} 
\affiliation{Department of Physics, University of Basel, 4056 Basel, Switzerland} 
\alsoaffiliation{Swiss Nanoscience Institute, University of Basel, 4056 Basel, Switzerland}
\email{floris.braakman@unibas.ch}

\date{\today}
\begin{document}
\begin{abstract}
Focused-electron-beam-induced deposition is a promising technique for patterning nanomagnets for spin qubit control in a single step. We fabricate cobalt nanomagnets in such a process, obtaining cobalt contents and saturation magnetizations comparable to or higher than those typically obtained using electron-beam lithography. We characterize the nanomagnets using transmission electron microscopy and image their stray magnetic field using scanning NV magnetometry, finding good agreement with micromagnetic simulations. The magnetometry reveals the presence of magnetic domains and halo side-deposits, which are common for this fabrication technique. Finally, we estimate dephasing times for electron spin qubits in the presence of disordered stray fields due to these side-deposits. 
\end{abstract}

\maketitle

Nanomagnets with precisely defined geometries are of interest for a variety of applications, including magnetic resonance force microscopy\cite{sidlesMagneticResonanceForce1995,degenNanoscaleMagneticResonance2009}, as mediating elements between spins and mechanical degrees of freedom\cite{rablQuantumSpinTransducer2010,gieselerSingleSpinMagnetomechanicsLevitated2020,rosenfeldEfficientEntanglementSpin2021}, magnetic memories\cite{slonczewskiCurrentdrivenExcitationMagnetic1996}, and for the implementation of quantum logic with spin-based qubits\cite{pioro-ladriereElectricallyDrivenSingleelectron2008,tokuraCoherentSingleElectron2006} such as electron spins confined in quantum dots.

Such electron spin qubits can be controlled and manipulated using high-frequency voltages applied to metallic gates\cite{burkardSemiconductorSpinQubits2021}, and selective spin rotation can be implemented by periodically displacing the electron wave function inside a magnetic field gradient resulting from a nearby nanomagnet \cite{pioro-ladriereElectricallyDrivenSingleelectron2008}. 
Recent experiments have shown successful operation in fault-tolerant regimes with gate fidelities above the required thresholds\cite{millsTwoqubitSiliconQuantum2022,noiriFastUniversalQuantum2022,xueQuantumLogicSpin2022}.
Realizing fast spin rotation, while at the same time keeping dephasing and relaxation rates acceptably low, requires precise engineering of strong magnetic field gradients. This places stringent constraints on the geometry, relative location and alignment, and magnetic properties of the used nanomagnets\cite{yonedaRobustMicromagnetDesign2015,neumannSimulationMicromagnetStrayfield2015}.  Furthermore, when scaling up to larger qubit arrays, the variability between a large number of individual nanomagnets will need to be characterized and minimized. Spatial characterization of nanomagnet stray fields is therefore important in order to facilitate qubit device fabrication, precise positioning of quantum dots relative to the nanomagnet, and to correctly assess and minimize qubit decohering mechanisms.

Typically, nanomagnets are patterned using a multi-step procedure, involving resist-coating, electron-beam lithography, metallization, and lift-off. Such a procedure is prone to introducing impurities in the devices due to residual resist particles, as well as to introducing possible misalignment. Furthermore, such techniques are limited to fabrication of 2D patterns.\\

Here, we use focused-electron-beam-induced deposition (FEBID) to pattern Co nanomagnets in a single step\cite{teresaReviewMagneticNanostructures2016}. FEBID is an appealing technique for the fabrication of nanomagnets integrated in qubit devices, since it generates no impurities in the form of residual resist, eases fabrication due to its single-step nature, and allows for the fabrication of 3D geometries\cite{magenFocusedElectronBeamEngineering3D2021}, opening up new ways of engineering magnetic gradients optimized for spin qubit control. FEBID of Co has been demonstrated as a reliable technique for growing highly magnetic nanostructures, reaching Co content of up to $\sim$96 atomic percent of bulk values\cite{fernandez-pachecoMagnetotransportPropertiesHighquality2009, teresaReviewMagneticNanostructures2016}. FEBID also allows for patterning with lateral resolution in the nm range\cite{vankouwenFocusedElectronBeamInducedDeposition2009}, approaching the intrinsic limit of the process imposed by the electron beam diameter\cite{silvis-cividjianSpatialResolutionLimits2005,salvador-porrocheHighlyefficientGrowthCobalt2021}. For Co nanostructures, lateral resolutions of below \SI{30}{\nano\meter} have thus far been achieved\cite{serrano-ramonUltrasmallFunctionalFerromagnetic2011}. 

\begin{figure*}[t]
    \includegraphics[width=0.96\textwidth]{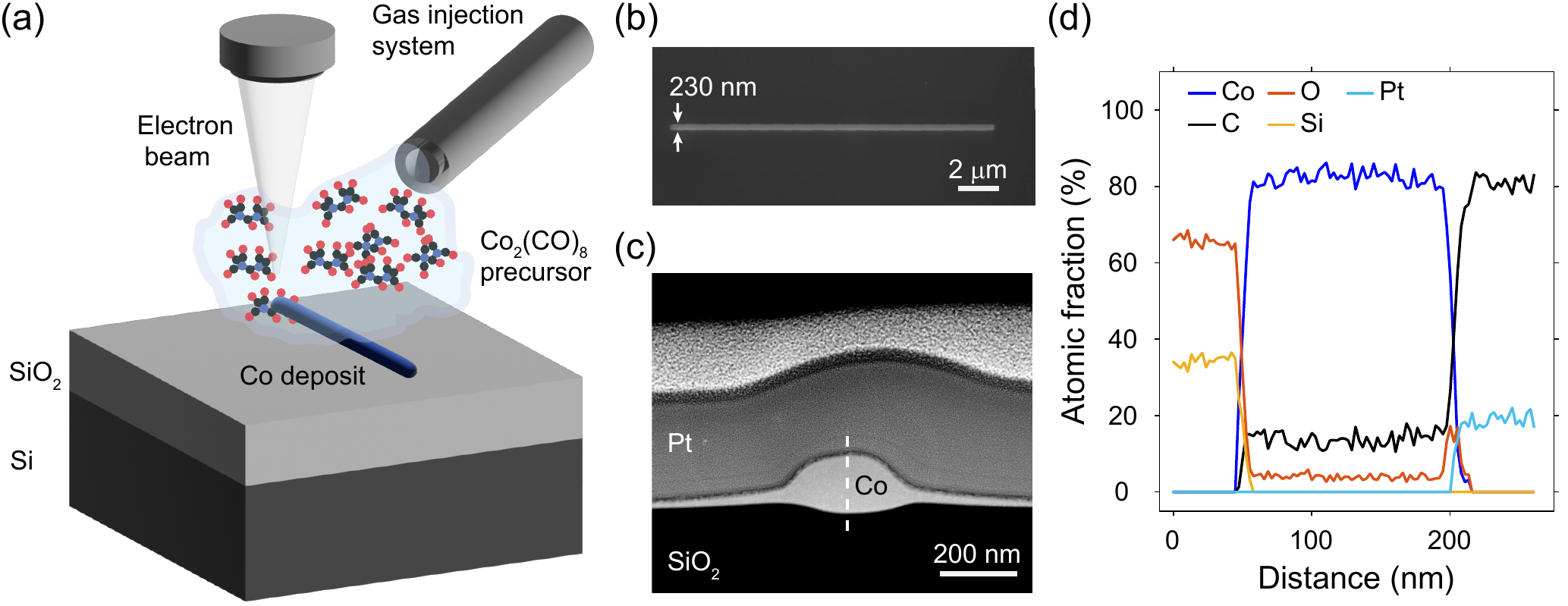}
    \caption{(a) Illustration describing FEBID patterning of Co nanomagnet. (b) SEM image of the Co nanomagnet studied here. (c) HAADF-STEM image of a cross-section of a Co NW nanomagnet, similar to the magnets studied here. (d) EDS analysis along linecut shown in (c).}
    \label{fig1}
\end{figure*}

We characterize the properties of FEBID nanomagnets using high-angle annular dark-field scanning transmission electron microscopy (HAADF-STEM), high-resolution energy dispersive spectroscopy (EDS) analysis, and atomic force microscopy (AFM). Next, We use scanning NV magnetometry (SNVM)\cite{rondinMagnetometryNitrogenvacancyDefects2014} to image the magnetic stray field of the Co deposits, both at externally applied magnetic field sufficiently high to achieve magnetization saturation, and at zero field. We find good agreement of our measurements with micromagnetic simulations. 
From our SNVM measurements of the disordered magnetic stray field of unintended deposits surrounding the nanomagnet, we estimate spin qubit dephasing times in the presence of charge noise. \\

The sketch in Fig.\ref{fig1}a illustrates the working principle of our FEBID\cite{randolphFocusedNanoscaleElectronBeamInduced2006,utkeGasassistedFocusedElectron2008,huthFocusedElectronBeam2012,utkeNanofabricationUsingFocused2012} fabrication technique. First, a precursor molecule containing cobalt, $\textrm{Co}_\textrm{2}\textrm{(CO)}_\textrm{8}$, is introduced inside a chamber pumped to high vacuum. Irradiating the precursor with an electron beam causes it to decompose, leaving Co deposits on a nearby sample substrate\cite{teresaReviewMagneticNanostructures2016}. By directing the beam using a scanning electron microscope, this technique can be used to, in a single step, pattern Co nanodeposits with high resolution\cite{serrano-ramonUltrasmallFunctionalFerromagnetic2011}. 

We use a Thermo Fisher FEI Helios 650 NanoLab FIB/SEM, fitted with a $\textrm{Co}_\textrm{2}\textrm{(CO)}_\textrm{8}$ gas injection system. The nanomagnets are patterned on the top surface of a Si substrate covered with \SI{290}{\nano\meter} of thermally grown SiO$_\textrm{2}$. We fabricate the nanomagnets with a nanowire (NW) shape, in order to obtain a magnetic configuration with a single magnetic easy axis, enabling simple alignment of our scanning probe and straightforward comparison to simulations. To achieve high Co content and high lateral resolution, we used the following FEBID parameters\cite{teresaReviewMagneticNanostructures2016}: an acceleration voltage of \SI{10}{\kilo\volt}, a beam current of \SI{3.2}{\nano\ampere}, a dwell time of \SI{1}{\micro\second}, and a precursor flux corresponding to a vacuum chamber pressure of \SI{4e-6}{\milli\bar}. Using these settings, we have achieved patterning Co structures with lateral widths down to \SI{50}{\nano\meter} and heights down to \SI{15}{nm}, as measured via AFM (See Supporting Information).\\

After FEBID fabrication, we use scanning and transmission electron microscopy (SEM and TEM) to characterize the geometry and composition of representative nanomagnets. Fig.\ref{fig1}b shows an SEM top-view image of a Co NW deposit (dimensions: \SI{14.4}{\micro\meter} length, \SI{230}{\nano\meter} width, and \SI{130}{\nano\meter} height) and Fig.\ref{fig1}c shows an HAADF-STEM image of a cross-section of such a deposit. In Fig.\ref{fig1}c, the rounded cross-section of the Co NW can be discerned, as well as "halo"  side-deposits of nm thickness extending laterally for several microns. EDS mapping along the linecut indicated in Fig.\ref{fig1}c reveals a composition consisting mostly of Co (82 $\pm$ 2.5 $\%$), with additional smaller amounts of C (14 $\pm$ 2.5 $\%$) and O (4 $\pm$ 2.5 $\%$) (see Fig.\ref{fig1}d). We find that this composition is rather uniform throughout the deposit, including similar proportions in the halo side-deposits (see Supporting Information for additional EDS data).\\ 

\begin{figure*}[t]
    \includegraphics[width=0.96\textwidth]{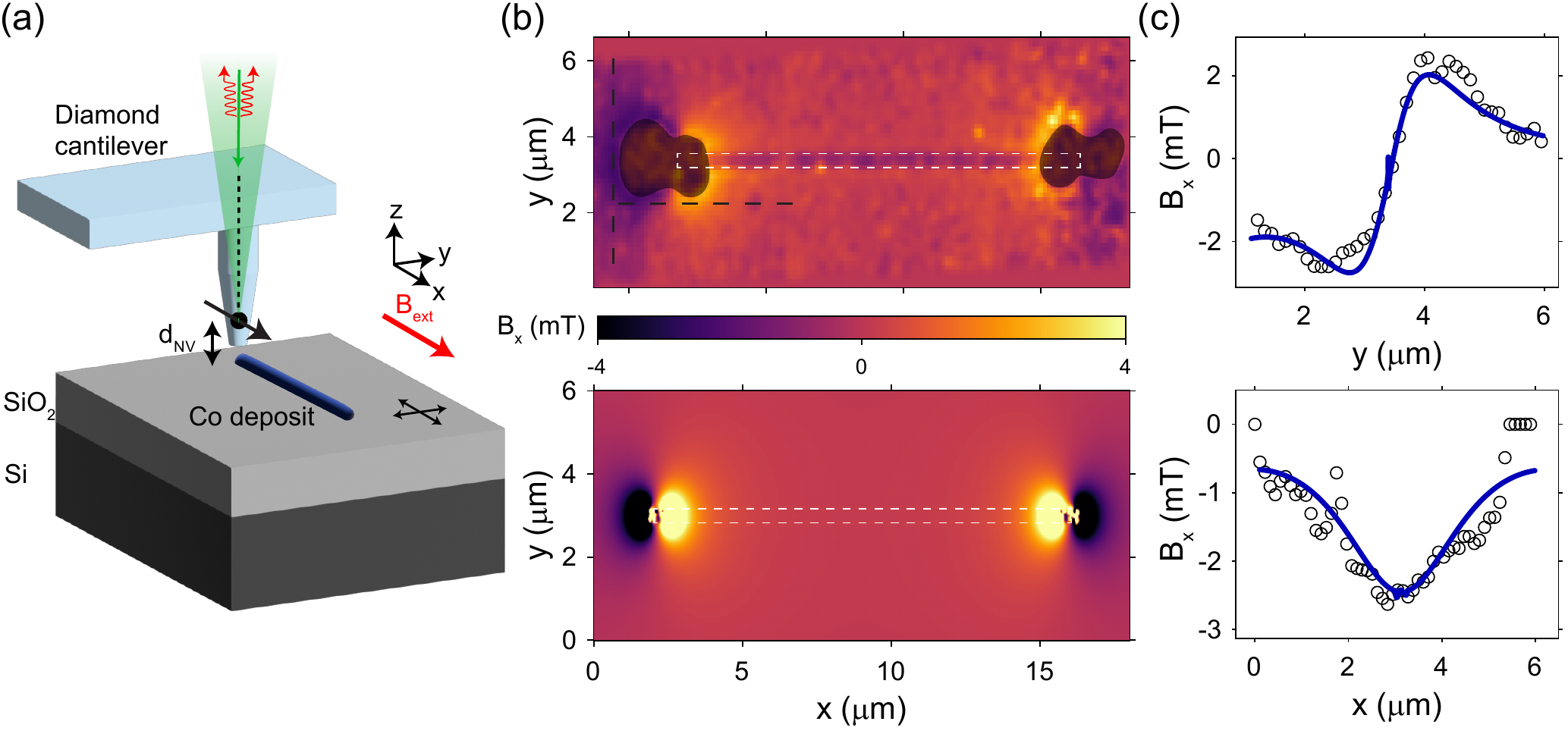}
    \caption{(a) Illustration of the SNVM setup. (b) Magnetic stray field produced by a Co NW, with $B_{ext}=$202.5 mT. Top panel: SNVM data. Data has been smoothened using a Gaussian filter with $\sigma$ below pixel size. Regions over the ends of the NW have been blacked out, since here the SNVM measurements are unreliable (see main text). Bottom panel: simulation taken at a height 30 nm above the bottom surface of the magnet geometry. (c) Horizontal (top panel) and vertical (bottom panel) linecuts of the SNVM data, at the positions indicated by black lines in (b). Solid blue lines are corresponding linecuts taken from the simulated data in (b).}
    \label{fig2}
\end{figure*}

The halo is commonly deposited as a side-effect in FEBID, produced through precursor dissociation by secondary electrons scattering off the substrate and the pattern that is being grown\cite{gavagninMagneticForceMicroscopy2014,shawravHighlyConductivePure2016}. Such halo deposits are typically undesirable and various approaches can be used to mitigate their formation. The amount of halo and its composition can vary depending on the deposition parameters, in particular the exact amount of precursor gas present in the chamber. Furthermore, by performing FEBID at low temperatures, halo effects can potentially be reduced. Finally, the halo can in principle be removed by means of argon ion milling (see Supporting Information), although at the same time a layer of the intended deposited structure and surrounding substrate may be removed and charge defects may be introduced into the device.\\

Compared to other scanning probe magnetometry techniques\cite{marchioriNanoscaleMagneticField2021}, such as scanning superconducting quantum interference device (SQUID) magnetometry and magnetic force microscopy (MFM), SNVM\cite{degenScanningMagneticField2008,degenNanoscaleMagneticResonance2009,schirhaglNitrogenVacancyCentersDiamond2014} offers several advantages which make it suitable for our use. Of particular relevance for our application is the high spatial resolution that can be achieved with SNVM, which can reach \SIrange{15}{25}{\nano\meter}\cite{changNanoscaleImagingCurrent2017,ariyaratneNanoscaleElectricalConductivity2018,marchioriNanoscaleMagneticField2021}, making it possible to image magnetic fields and currents at length scales relevant for spin qubit devices. Also, SNVM yields quantitative measurements of the magnetic fields as the Zeeman energy of a single NV-center defect can be probed directly. Furthermore, due to its high magnetic field sensitivity on the order of $\mu \textrm{T}/\sqrt{\textrm{Hz}}$, SNVM allows imaging the weak fields associated with nanoscale magnetic domains \cite{celanoProbingMagneticDefects2021} and other spatially inhomogeneous magnetic stray fields, making it a useful tool to study the magnetization properties of FEBID Co halo structures and their impact on spin qubit performance.\\

Fig.\ref{fig2}a illustrates the SNVM setup employed here: a commercial system (Qnami ProteusQ) operating under ambient conditions. We use a diamond cantilever (Qnami Quantilever MX) hosting a single negatively charged NV center embedded inside its protruding tip. The cantilever is attached to a quartz Akiyama tuning fork, allowing for frequency-modulated AFM. For our measurements we use diamond tips hosting an NV center with a spin quantization axis oriented parallel to the principal axis of the NW magnet, i.e. along the $x$-axis as defined in Fig.\ref{fig2}a. This type of diamond tips are fabricated from 110 diamond blankets\cite{maertzVectorMagneticField2010}. We estimate the distance $d_{NV}$ of the NV center to the apex of the diamond tips to be \SIrange[]{30}{50}{\nano\meter}\cite{celanoProbingMagneticDefects2021}, and corresponding best achievable lateral spatial resolutions of $0.86\cdot d_{NV}$. During the measurements, an external magnetic field $B_{ext}$ is applied along the NV quantization axis. This direction coincides with the principal NW axis and its easy magnetic axis. To perform magnetometry, we employ measurements of optically detected electron spin resonance as well as of fluorescence. See e.g. Celano et al.\cite{celanoProbingMagneticDefects2021} for a more in-depth description of the SNVM setup and measurement techniques used here.\\

Fig.\ref{fig2}b shows an SNVM scan of a Co NW nanomagnet, taken with $B_{ext}$ = \SI{202.5}{\milli\tesla}, which falls within the typical operating range of spin qubits. The scan is taken with a tip-sample distance < \SI{5}{\nano\meter}, and consequently the magnetometry measurements are taken at a distance $\sim d_{NV}$ from the sample surface. The scan shows the $x$-component of the magnetic stray field of the nanowire-shaped magnet, revealing a pole at each end of the magnet. At this value of $B_{ext}$, the nanomagnet is almost fully saturated along its magnetic easy axis. The associated stray field profile features large regions surrounding the nanomagnet where field components transverse to the quantization axis of the NV center are  small. In these regions, relatively little quenching of NV fluorescence\cite{rondinMagnetometryNitrogenvacancyDefects2014} occurs and it is straightforward to reconstruct the $x$-components of the stray field from the SNVM measurements. Even so, we blacked out regions in Fig.\ref{fig2}b where we could not reliably track the Zeeman splitting of the NV center. This can occur when the magnetic stray field is too large, transverse components are too large, or when the optical read-out signal is quenched. Especially at the ends of the NW, we expect strong out-of-plane stray field components. These out-of-plane fields are transverse to the NV axis and lead to a quenching of the NV signal\cite{rondinMagnetometryNitrogenvacancyDefects2014}.

We compare the SNVM measurement with finite-element simulations of the $x$-component of the stray field in the same area around the NW (see Fig.\ref{fig2}b, lower panel), using the software package MuMax3.\cite{vansteenkisteDesignVerificationMuMax32014,exlLaBonteMethodRevisited2014} Here, we simulate the stray field of a rectangular Co box geometry with a width of \SI{250}{\nano\meter}, height of \SI{130}{\nano\meter}, and a length of \SI{14.4}{\micro\meter}. We use a typical value of the exchange constant for Co, $A_{ex} = 14\cdot 10^{-12}\textrm{J/m}$, and a 5x5x5 nm$^3$ cell size. Using this model, we obtain a $B_{x}$ stray field profile that qualitatively agrees well with the experiment, as shown in Fig.\ref{fig2}b. Fig.\ref{fig2}c shows plots of vertical and horizontal linecuts taken at the corresponding lines shown in Fig.\ref{fig2}b. We find best agreement between simulation and experiment when we use a saturation magnetization of $M_{s} = 1.2\cdot 10^{6} \textrm{A/m}$ in the simulation. Such a saturation magnetization corresponds to 85\% of the bulk value, agreeing well with the atomic fraction of Co measured in our deposit (Fig.\ref{fig1}d). We note that exactly aligning the simulation with the experimental data in the $xy$-plane is to some degree hindered by imperfect knowledge of the precise location of the NV center inside the scanning tip, as well as by the pixel size of the scan. Some deviations between simulation and experiment may also result from the fact that in the simulation we do not take the rounded shape of the NW or the halo into account.\\

\begin{figure}[t]
    \includegraphics[width=0.70\textwidth]{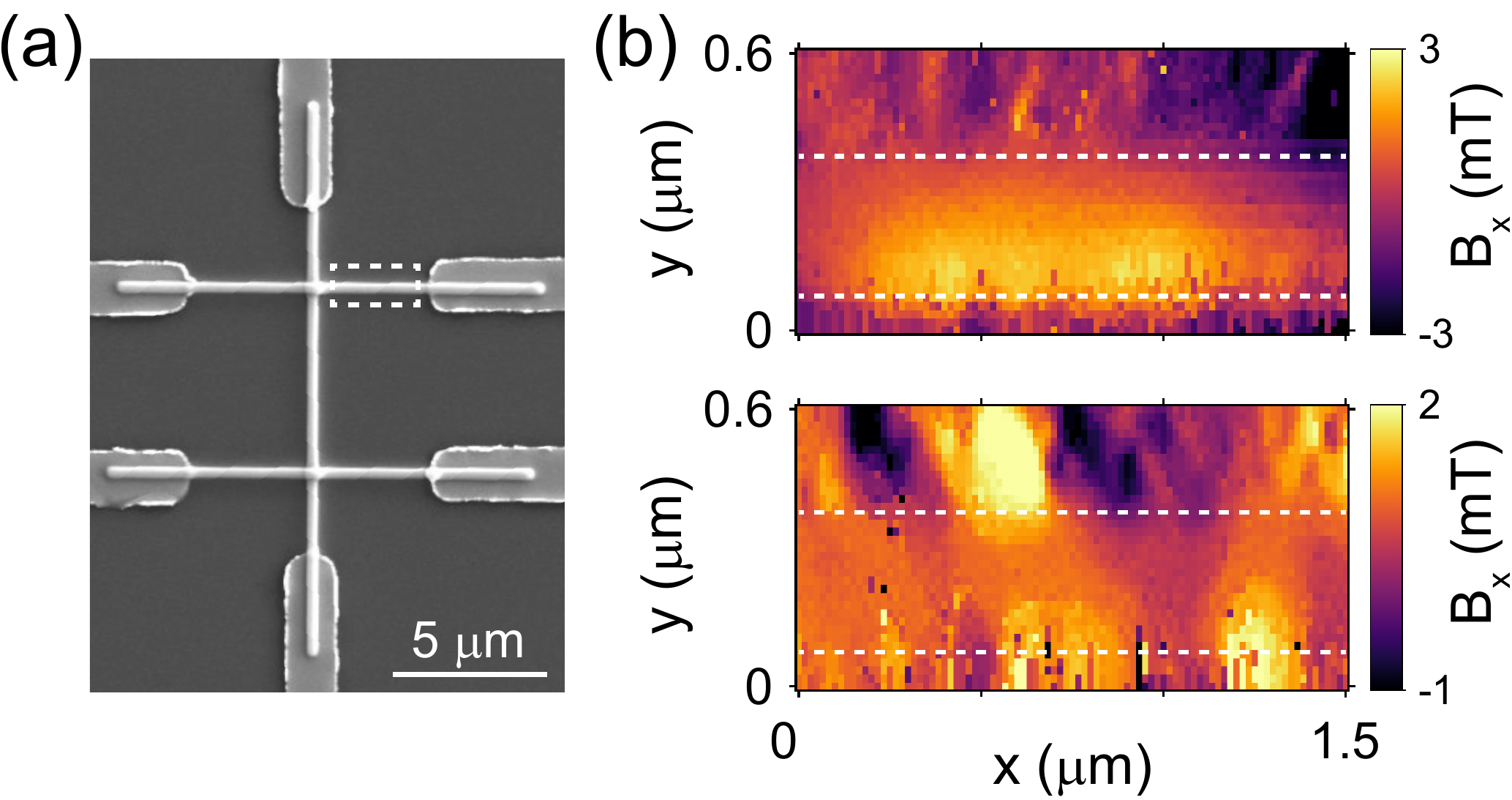}
    \caption{FEBID Co Hall bar. (a) SEM image of Co Hall bar device patterned through FEBID. Contacts are Ti/Pd, fabricated using electron-beam lithography and lift-off procedure. (b) SNVM maps of Co NW taken at $B_{ext}$ = 240 mT (upper panel), and $B_{ext}$ = 13 mT (lower panel). White dashed lines indicate location of NW.}
    \label{fig3}
\end{figure}

We further investigate the presence of magnetic structures of characteristic sizes of \SIrange{50}{200}{\nano\meter}. Such dimensions are of the same order of magnitude as the length scales relevant for spin qubit devices, such as the typical dimensions of quantum dots, confinement gate electrodes, nanomagnets, and coupling elements\cite{burkardSemiconductorSpinQubits2021}. Moreover, also tunneling lengths and typical wave function displacements are of a similar order of magnitude. Of particular relevance for spin qubits are unintended variations of the magnetic stray field on short length scales. In the presence of small displacements of the electron wave function, such variations can translate to magnetic noise, which can limit qubit decoherence~\cite{hansonSpinsFewelectronQuantum2007,neumannSimulationMicromagnetStrayfield2015}.

In the Co nanomagnet devices shown here, we find small magnetic structures in the form of magnetic domains inside the NW at low $B_{ext}$, as well as grain-like stray fields produced by the halo deposits surrounding the NW. We investigate these smaller structures in a Hall bar device consisting of 3 crossing Co NWs fabricated through FEBID using similar parameters as before, see Fig.\ref{fig3}a. Fig.\ref{fig3}b shows SNVM scans of a part of one of the Co NWs, in the region delineated in Fig.\ref{fig3}a. In the upper panel of Fig.\ref{fig3}b, SNVM data taken at $B_{ext}$ = \SI{240}{\milli\tesla} is shown, at which field the magnetization of the horizontal NW section is saturated, resulting in a homogeneous stray field above the NW. In the lower panel in Fig.\ref{fig3}b, SNVM of the same section is shown at $B_{ext}$ = \SI{13}{\milli\tesla}. In this case, multiple domains of characteristic size of several hundred nanometer can be discerned in the observed stray field of the magnet including its halo. \\

\begin{figure*}[t]
    \includegraphics[width=0.95\textwidth]{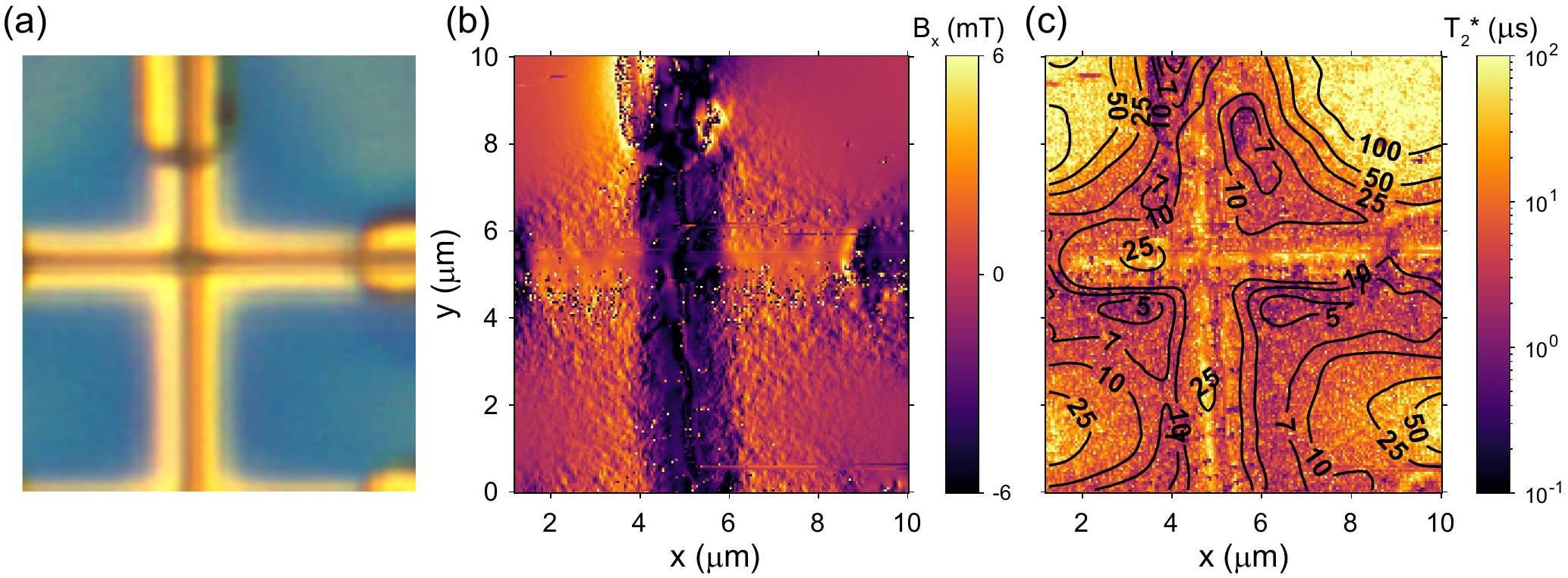}
    \caption{(a) Optical micrograph of Co Hall bar, with halo distinguishable as dark shape. (b) SNVM map of same area as in (a), taken at $B_{ext}$ = 13 mT. Dark vertical central line and areas at contacts feature significant field components transverse to the NV quantization axis, preventing straightforward determination of $B_x$ in these areas. (c) Map of estimated $T_2^*(x,y)$ for the same area, assuming spin qubit rms displacement amplitude of \SI{10}{\pico\meter}.}
    \label{fig4}
\end{figure*}

Next, we use SNVM to study the halo side-deposits in more detail. Figs.\ref{fig4}a, and b show optical microscopy and SNVM images, respectively, of a Co Hall bar structure patterned via FEBID, which exhibits a significant halo side deposit. As can be seen, the halo can be distinguished as a dark shade of inhomogeneous shape in the optical microscopy image. The SNVM image of the same area further reveals that the halo presents a magnetic stray field of grainy composition, see Fig.\ref{fig4}b (see Supporting Information for a magnified figure). The grainy pattern follows the same shape as the dark shade discernible in Fig.\ref{fig4}a: it surrounds the intended deposit and becomes smooth further away from the deposit. We investigate the size distribution of the grainy structures, using a segmentation analysis (Gwyddion) on a subset of the data shown in Fig.\ref{fig4}b. We find a typical equivalent square side $a_{eq}$ of roughly \SI{100}{\nano\meter}, larger than the pixel size of 50x50 nm of the scan. Furthermore, we find associated stray field fluctuations of up to 3 mT. \\

We estimate the effect of such magnetic stray field fluctuations on the dephasing of electron spin qubits when placed inside the stray field of Fig.\ref{fig4}b. Specifically, we consider dephasing as a result of spin qubit displacements inside the halo stray field due to charge noise. Typical rms displacement amplitudes of QD electron spin qubits in this scenario are \SIrange{1}{10}{\pico\meter}\cite{neumannSimulationMicromagnetStrayfield2015,kawakamiElectricalControlLonglived2014}, along $x$ and $y$. Out-of-plane displacements are typically negligible for quantum well or MOS quantum dots, since the confinement potential in this direction is much larger than in the $xy$-plane.  Since such displacements are orders of magnitude smaller than the grain size of the halo stray field, we can restrict our analysis to using the first derivative at each point, and neglect high-order derivatives of the stray field. Taking the spin qubit quantization axis to be parallel to $x$, the stray field derivatives that are relevant for dephasing are therefore $dB_x/dx$ and $dB_x/dy$. By differentiating the scan of Fig.\ref{fig4}b with respect to $x$ and $y$, we find that both $dB_x/dx$ and $dB_x/dy$ are largest for positions near the intended Co Hall bar deposit (top left and right corners, slightly above bottom left and right corners of plot in Fig.\ref{fig4}b), but do not exceed 425$\mu$T/nm at any point of the scan. \\

Using these derivatives of the stray field, we can estimate the inhomogeneous dephasing time $T_2^*$ of a spin qubit placed inside the grainy stray field induced by the halo. For each point $(x,y)$ of the scan of Fig.\ref{fig4}b, we calculate $T_2^*(x,y)$ using $T_2^* = \sum\limits_{i}(2\pi\sqrt{2}\cdot\mu_B/\hbar\cdot dB_x/di\cdot \Delta i)^{-1}$, with $i\in{x,y}$. Here we use $\Delta x = \Delta y =$ \SI{10}{\pico\meter}\cite{neumannSimulationMicromagnetStrayfield2015}, an electron spin Land\'{e} g-factor of 2, and we assume a quasi-static $1/f$-like spectral density of the charge noise\cite{nakajimaCoherenceDrivenElectron2020a}. Fig.\ref{fig4}c shows the corresponding map of $T_2^*(x,y)$. In this case, $T_2^*(x,y)$ decreases from several hundreds of $\mu$s on the bottom-right side of the scan, where almost no halo is present, to roughly \SI{10}{\micro\second} in the top-left of the scan, where the halo is most intense. We find that $T_2^*$ exceeds \SI{1}{\micro\second} for each point of the scan. Note that the contours in the plot of Fig.\ref{fig4}c have been obtained by smoothing the data. While these contours indicate the trend of decreasing $T_2^*$ when approaching the Co deposit, the grainy pattern visible in the colorplot of Fig.\ref{fig4}c originates from the disordered halo stray field, and hence should not be ignored.\\

The estimated $T_2^*(x,y)$ shown in Fig.\ref{fig4}e are on par with those found for various kinds of high-quality spin qubits in Si- and Ge-based quantum dots\cite{stanoReviewPerformanceMetrics2022}, indicating that the spatially inhomogeneous stray fields of the halo side-deposits need not limit coherence more than other factors, such as charge noise in the presence of strong intended field gradients or spin-orbit coupling, and hyperfine interactions. In future work, we aim to characterize also the time-dependent magnetic noise originating from the halo and evaluate its impact on spin qubits. from Our TEM and SNVM characterization show that our FEBID structures have Co content and saturation magnetization comparable or higher than what is typically obtained using Co evaporation and standard electron beam lithography patterning. Moreover, past results have shown that depositions of Co content in excess of 95 atomic percent can be obtained using FEBID. Such high Co contents, in combination with the ability of FEBID to produce 3D magnet geometries would enable further optimization of nanomagnets for spin qubit control. 

Finally, future research may target cryo-FEBID for the patterning of magnetic nanostructures on sensitive spin qubit devices, since it allows to pattern deposits with electron doses of order $10^3$ $\mu$C/cm$^2$, which is $\sim10^4$ times less than needed for FEBID at room temperature\cite{bresinDirectwrite3DNanolithography2013,deteresaComparisonFocusedElectron2019}, and similar to what is used in electron-beam exposure of resists. Hence, it can be expected that sample damage due to electron irradiation is comparable for cryo-FEBID and resist-based electron-beam lithography techniques.

\begin{acknowledgement}
We thank Prof. José María De Teresa, Prof. Patrick Maletinsky, and Dr. Monica Sch\"onenberger for useful discussions and assisting with the AFM measurements. Calculations were performed at sciCORE (\url{http://scicore.unibas.ch}) scientific computing center at University of Basel. We acknowledge funding from the Swiss National Science Foundation via NCCR SPIN as well as Project grant 200020\textunderscore 207933.
\end{acknowledgement}

\nocite{*}

\bibliography{NVpaperbetterbibtex}
\end{document}